\documentclass[aps,prd,superscriptaddress,reprint,nofootinbib]{revtex4-1}
\usepackage{amsthm,amsmath,physics,mathtools,amssymb}

\usepackage{autobreak}

\usepackage{hyperref}
\hypersetup{
    colorlinks=true,
    linkcolor=blue,
    filecolor=magenta,      
    urlcolor=cyan,
}

\usepackage{cleveref}

\usepackage{tabularx}

\usepackage{array}
\makeatletter
\newcommand{\thickhline}{%
    \noalign {\ifnum 0=`}\fi \hrule height 1.5pt
    \futurelet \reserved@a \@xhline
}

\usepackage{CJKutf8}

\usepackage{graphicx}
\graphicspath{ {images/} }

\begin{document}

\begin{CJK*}{UTF8}{gbsn}
\title{Causally neutral quantum physics}
\author{Ding Jia (贾丁)}
\email{ding.jia@uwaterloo.ca}
\affiliation{Department of Applied Mathematics, University of Waterloo, Waterloo, Ontario, N2L 3G1, Canada}
\affiliation{Perimeter Institute for Theoretical Physics, Waterloo, Ontario, N2L 2Y5, Canada}

\begin{abstract}
In fundamental theories that accounts for quantum gravitational effects, the spacetime causal structure is expected to be quantum uncertain. Previous studies of quantum causal structure focused on finite-dimensional systems. Here we present an algebraic framework that incorporates both finite- and infinite-dimensional systems including quantum fields. Thanks to the absence of a definite spacetime causal structure, Lagrangian quantum field theories can be studied on a quantum superposition of spacetimes with a point identification structure.
\end{abstract}

\maketitle
\end{CJK*}

\section{Introduction}

This work is guided by the \textbf{principle of causal neutrality}:
\begin{quote}
Fundamental concepts and laws of physics should be stated without assuming a definite spacetime causal structure.
\end{quote}
In general relativity, gravity is nine tenth causal structure (the other one tenth is the conformal factor) \cite{hawking1973large, hawking_new_1976}. When gravity is subject to quantum fluctuations, the spacetime causal structure most likely becomes indefinite. This rationale motivates the above principle.

Many of our current concepts and theories of physics need to be generalized or modified if the principle of causal neutrality holds. As an example for a concept, entanglement is traditionally considered for spacelike separated systems. In the presence of indefinite causal structure two systems cannot be said to be definitely spacelike separated, so the meaning of entanglement needs to be clarified. As an example for a theory, traditional quantum field theory is based on spacetime manifolds with definite causal structure, which is used to phrase the basic axioms such as microcausality. In the presence of indefinite causal structure spacetime manifolds with definite causal structure cannot be retained, so the framework of quantum field theory needs either to be generalized or discarded.

To our knowledge Hardy first promoted accommodating indefinite causal structure as a central feature for the quantum theory of gravity and formulated a framework for general probabilistic theories that does not assume definite spacetime causal structure \cite{hardy2005probability, hardy2007towards}. 
More recently, several frameworks accommodating indefinite causal structure for quantum theory were proposed using tools from quantum information theory, e.g., \cite{chiribella2013quantum, *chiribella2009quantum, perinotti2017causal, bisio2018axiomatic, oreshkov2012quantum, araujo2015witnessing, oreshkov2016causal, oreshkov2015operational, oreshkov2016operational, giacomini2016indefinite, silva2017connecting}. Various information processing protocols taking advantage of indefinite causal structure were found \cite{chiribella2012perfect, araujo2014computational, feix2015quantum, ried2015quantum, guerin2016exponential, ebler2018enhanced}, and the experimental realization of processes with indefinite causal structure had become a very active area of research \cite{procopio2015experimental, maclean2017quantum, rubino2017experimental, Goswami2018}.

Here we present an algebraic framework for causally neutral quantum physics incorporating finite- and infinite-dimensional systems, including quantum fields. This framework provides a starting point to generalize quantum field theories such as QED to include effects of indefinite causal structure, and to study the impacts of indefinite causal structure on field entanglement, field regularization, the Unruh and Hawking radiation and other topics for which both quantum fields and causal structure are relevant. 

The following is a list of characteristic features of the new framework.
\begin{enumerate}
\item Incorporates indefinite causal structure for finite-dimensional systems
\cite{hardy2005probability, hardy2007towards, chiribella2013quantum, *chiribella2009quantum, perinotti2017causal, bisio2018axiomatic, oreshkov2012quantum, araujo2015witnessing, oreshkov2016causal, oreshkov2015operational, oreshkov2016operational, silva2017connecting} and infinite-dimensional systems, \cite{giacomini2016indefinite}\footnote{\cite{giacomini2016indefinite} incorporates continuous-variable systems, but not quantum fields.} including quantum fields. 
\item Does not assume a spacetime manifold as a basic ingredient. \cite{oeckl2003general, oeckl2013positive, raasakka2017spacetime, hardy2018construction}
\footnote{All the frameworks cited under the first item (using tools of quantum information theory) do not require a spacetime manifold. Oeckl's general boundary framework \cite{oeckl2003general, oeckl2013positive} refers to smooth manifolds but not the metric. Works of background-independent quantum theory following the boundary approach also do not refer to a spacetime manifold. See, e.g, \cite{rovelli2004quantum} and references therein.}
\item Assumes abstract algebras and functionals, but not Hilbert spaces \footnote{Not presuming Hilbert spaces is an advantage for studying the foundations of QFT, as stressed in the algebraic approach \cite{haag1996local}. In \Cref{sec:hs} we show a construction of Hilbert spaces analogous to the GNS construction.} as basic ingredients. \cite{haag1964algebraic, haag1996local, raasakka2017spacetime}
\item Uses two copies of the algebra/observable set. \cite{Schwinger1961Brownian, Keldysh1965DIAGRAM, Feynman1963Theory, hartle1993spacetime, sorkin1994quantum, silva2017connecting, Cotler2018Superdensity}
\end{enumerate}
While the cited works share the individual features, to our knowledge the present framework is unique is possessing all the features. The main structures of the framework are summarized in \Cref{tab:compare} through a comparison with traditional quantum physics in the algebraic formulation.

\begin{table*}[t]
\centering
\caption{The present framework compared with the traditional framework}
\label{tab:compare}
\begin{tabular}{|p{.45\textwidth}|p{.45\textwidth}|}
\thickhline
               Traditional algebraic quantum physics                                   & Causally neutral quantum physics 
               \\ \thickhline
$(\mathcal{M},g_{ab})$ - spacetime manifold              & No \textit{a priori} reference to a spacetime manifold
\\ \thickhline
$\mathcal{A}$ - field/observable algebra                &  $\mathfrak{A}=\star_{i\in I}\mathfrak{A}_i$ - free product algebra 
\\ \hline causal structure relevant for $\mathcal{A}$, & causal structure irrelevant for $\mathfrak{A}$,
\\e.g., $[\phi_i,\phi_j]=0$ for spacelike separation & $[\phi_i,\phi_j]=0$ for $\phi_i\in \mathfrak{A}_i,\phi_j\in \mathfrak{A}_j$, as long as $i\ne j$
\\ \thickhline
 $\omega:\mathcal{A}\rightarrow \mathbb{C}$ - states                        & $\omega:\mathfrak{A}\times \mathfrak{A}\rightarrow \mathbb{C}$ - generalized states
                \\ \hline
linear functional               & bilinear functional
               \\ \hline
$\omega(a^*a)\ge 0$                & $\omega(a^*,a)\ge 0$
\\ \hline
$\omega(e)=1$                & $\omega(e,e)=1$
\\ \hline
causal structure irrelevant for $\omega$    & (indefinite) causal structure relevant for $\omega$ 
\\ \thickhline
\end{tabular}
\end{table*}

\section{The free product algebra}

In traditional quantum physics in the algebraic formulation, the algebra carries information about the definite causal structure. For instance, by the microcausality axiom, if $\phi_i$ and $\phi_j$ are from causally disconnected regions, then $[\phi_i,\phi_j]=0$. Consequently, $[\phi_i,\phi_j]\ne 0$ would imply that the two regions are definitely causally connected.

In the new framework the causal structure is \textit{not} reflected in the algebraic structure. We assume that there is a family of algebras to start with. These algebras are referred to as ``factor'' algebras because we will form a ``product'' algebra out of them. To avoid committing to a definite causal structure, we do not require that the algebras be attached to a background spacetime. 
In the global product algebra of the new framework, $[\phi_i,\phi_j]=0$ identically for $\phi_i$ and $\phi_j$ from different factor algebras  whatever the causal relation is. This global ``free product algebra'', adopted from Raasakka's spacetime-free quantum theory \cite{raasakka2017spacetime}, imposes no non-trivial algebraic relations on elements from different factor algebras. We first give a definition of the free product algebra through universal properties as the coproduct in the category of unital $*$-algebras. We then give a less abstract characterization of the free product algebra through its generators. 

Given a family $\{\mathfrak{A}_i\}_{i\in I}$ of unital $*$-algebras, their \textbf{free product algebra}  $\mathfrak{A}=\star_{i\in I}\mathfrak{A}_i$ with the unital *-homomorphisms $\psi_i:\mathfrak{A}_i\rightarrow\mathfrak{A}$ is the unique unital *-algebra satisfying the following universal property. 
\begin{quote}
    Given any unital *-algebra $\mathfrak{B}$ and unital *-homomorphisms $\phi_i:\mathfrak{A}_i\rightarrow \mathfrak{B}$, there exists a unique unital *-homomorphism $\Phi:\mathfrak{A}\rightarrow \mathfrak{B}$ so that $\phi_i=\Phi\circ\psi_i$ 
\end{quote}

For two unital *-algebras $\mathfrak{A}_1$ and $\mathfrak{A}_2$, $\mathfrak{A}_1\star\mathfrak{A}_2$ is linearly generated by finite sequences of the form $x_1 x_2\cdots x_n$, where $x_k\in \mathfrak{A}_1$ or  $x_k\in \mathfrak{A}_2$ for all $k$. For two such sequences $x_1 x_2\cdots x_n,y_1 y_2\cdots y_n\in \mathfrak{A}_1\star\mathfrak{A}_2$, the product is simply the concatenation 
\begin{align}
(x_1 x_2\cdots x_n)\star(y_1 y_2\cdots y_n)=x_1 x_2\cdots x_n y_1 y_2\cdots y_n.
\end{align}
The *-operation of $\mathfrak{A}_1\star\mathfrak{A}_2$ is simply given by
\begin{align}
(x_1 x_2\cdots x_n)^*=x_n^* x_{n-1}^*\cdots x_1^*.
\end{align}
Two equivalence relations are imposed on $\mathfrak{A}=\mathfrak{A}_1\star\mathfrak{A}_2$. First, the unit elements of $\mathfrak{A}_1$, $\mathfrak{A}_2$ and $\mathfrak{A}$ are identified:
\begin{align}
e_{\mathfrak{A}_1}\sim e_{\mathfrak{A}_2}\sim e_{\mathfrak{A}}.
\end{align}
Second, if $x_k, x_{k+1}$ belong to the same $\mathfrak{A}_i$ and $x'=x_k x_{k+1}$, then
\begin{align}
x_1 x_2\cdots x_k x_{k+1}\cdots x_n \sim x_1 x_2\cdots x'\cdots x_n.
\end{align}

\section{The generalized states}

Given $\mathfrak{A}$ as a unital *-algebra (over $\mathbb{C}$), here taken to be a free product algebra, we define a \textbf{generalized state} as a bilinear functional $\omega:\mathfrak{A}\times \mathfrak{A}\rightarrow \mathbb{C}$ satisfying
\begin{align}
\omega(a^*,a)\ge 0,
\\
\omega(e,e)=1,
\end{align}
where $e$ is the unit element.
In traditional algebraic quantum physics \cite{haag1996local} a state $\omega':\mathcal{A}\rightarrow\mathbb{C}$ on a $*$-algebra or a $\mathbb{C}^*$-algebra $\mathcal{A}$ obeys the conditions $\omega'(a^*a)\ge 0$ and $\omega'(e)=1$. $\omega(a^*,a)\ge 0$ and $\omega(e,e)=1$ are analogues of these conditions.

The states $\omega$ are ``generalized'' because in contrast to traditional states they carry information about the dynamical correlations of the algebras. This is the major conceptual shift from traditional frameworks. It enables the new framework to incorporate more general correlations, such as those encoding indefinite causal structure. Before demonstrating this, we show how to express traditional correlation functions in the new framework.

\section{Traditional correlation functions}

Consider the traditional QFT correlation function
\begin{align}\label{eq:oqftcf}
\bra{\psi}\tilde{x} \tilde{y}\ket{\psi},
\end{align}
where $\ket{\psi}$ is a vector state in some Hilbert space, and the operators $\tilde{x},\tilde{y}$ belong to the traditional QFT algebra $\mathcal{A}$. For example, these can be field operators.\footnote{In the presentation below we work with unsmeared operators for simplicity, and omit writing out the smeared version of the expressions, which can be obtained in the standard way be supplying test functions.}

In traditional QFT the algebraic elements carry time evolution in themselves. The field operators are usually taken to be Heisenberg picture operators, so that $\tilde{x}=U^*_1 x U_1$, where $U_1$ is the time evolution unitary from the time of the state $\psi$ to the time of $\tilde{x}$, and $x$ is the corresponding Schr{\"o}dinger picture operator. Similarly $\tilde{y}=U^*_2 y U_2$. In terms of the ``untilde'' operators, (\ref{eq:oqftcf}) becomes
\begin{align}\label{eq:oqftcf2}
\bra{\psi}(U^*_1 x U_1) (U^*_2 y U_2)\ket{\psi}=\bra{\psi_1} x U_{1,2} y \ket{\psi_2},
\end{align}
where $\ket{\psi_1}=U_1\ket{\psi}$, $\ket{\psi_2}=U_2\ket{\psi}$, and $U_{1,2}=U_1 U^*_2$. 

In the new framework we use the algebras of the untilde elements. For the ``two-point'' correlation functions there are two original algebras $\mathfrak{A}_1$ and $\mathfrak{A}_2$. The free product $\mathfrak{A}=\mathfrak{A}_1 \star \mathfrak{A}_2$ is generated by terms of the form $a=x_ay_a\cdots z_a$, where $x_a,y_a,\cdots, z_a$ are elements of either $\mathfrak{A}_1$ or $\mathfrak{A}_2$. Define the generalized state by
\begin{align}\label{eq:otpcf}
&\omega:\mathfrak{A}\times \mathfrak{A}\rightarrow\mathbb{C},\nonumber\\ 
\omega(b,a)=\bra{\psi_2} &(v_b\cdots) U_{2,1} (u_b\cdots) 
\sum_{\ket{\psi_1}}\ketbra{\psi_1} \nonumber\\
& (x_a\cdots) U_{1,2} (y_a\cdots) \ket{\psi_2}.
\end{align}
Here we grouped elements of $a$ and $b$ according to their original factors. The factor $(x_a\cdots)$ collects all the elements of $a$ coming from $\mathfrak{A}_1$ in the order of their appearance, the factor $(y_a\cdots)$ collects all the elements of $a$ coming from $\mathfrak{A}_2$ in the order of their appearance, and similarly for $b=u_bv_b\cdots w_b$. (Each factor can in fact be reduced to a single element in $\mathfrak{A}_1$ or $\mathfrak{A}_2$.) The sum $\sum_{\ket{\psi_1}}$ is over an orthonormal basis of the state space. It is easy to check that $\omega$ is a generalized state according to the definition. The traditional QFT ``two-point'' correlation function (\ref{eq:oqftcf}) $\bra{\psi}\tilde{x} \tilde{y}\ket{\psi}=\bra{\psi}(U^*_1 x U_1) (U^*_2 y U_2)\ket{\psi}=\bra{\psi_1} x U_{1,2} y \ket{\psi_2}$ is recovered as $\omega(e,a)$ with $e$ as the unit element and $a=xy$. An $n$-point correlation function can be recovered analogously by introducing more entries such as $x$ and $y$, along with more $U$'s to connect the entries.

Note that the untilde elements of the algebra do not know about $U$, but the generalized state $\omega$ does. In other words, the generalized state $\omega$ carries the dynamical evolution.

\section{The quantum switch}

Freeing the algebra from the dynamics allows the framework to incorporate additional non-traditional processes. The quantum switch (due to the lack of a global time foliation) is an example that is unclear how to incorporate in traditional QFT frameworks but is straightforwardly described in the new framework.

The quantum switch expresses two operations in a ``superposition'' of causal order. It is widely studied for finite-dimensional information processing and multiple tasks had been found where it outperforms quantum circuits with definite causal structure (e.g., \cite{chiribella2013quantum, *chiribella2009quantum, chiribella2012perfect, araujo2014computational, feix2015quantum, guerin2016exponential, ebler2018enhanced}). A version of the quantum switch had also been conceived in connection with gravitational time dilation \cite{zych2017bell}. With some slight generalizations \footnote{Specifically, we allow for infinite-dimensional systems and for different evolutions (the $U^{(0)}$'s and $U^{(1)}$'s) for the $\ketbra{0}$ and $\ketbra{1}$ parts.}, the quantum switch can be described for infinite-dimensional systems including quantum fields as follows. 

The transition amplitude from $\ket{\psi}$ to $\ket{\phi}$ is given by 
\begin{align}\label{eq:qsta}
\bra{\phi}v \Big(&\ketbra{0}\otimes U_{v,x}^{(0)} x U_{x,y}^{(0)} y U_{y,u}^{(0)}+
\nonumber\\
&\ketbra{1}\otimes U_{v,y}^{(1)} y U_{y,x}^{(1)} x U_{x,u}^{(1)}\Big)u \ket{\psi}. 
\end{align}
The initial state $\psi=\psi'\otimes\psi''$ factors into a qubit $\psi'$ and the rest $\psi''$. When $\ket{\psi'}=\ket{0}$, $\ket{\psi''}$ goes through $y$ and then $x$.  When $\ket{\psi'}=\ket{1}$, $\ket{\psi''}$ goes through $x$ and then $y$. For a generic $\ket{\psi'}$, $\ket{\psi''}$ goes through $x$ and $y$ in a quantum superposition of different orders.

Denote the amplitude (\ref{eq:qsta}) by $\mathcal{A}(\ket{\phi},x,y,u,v)$. There are four original factor algebras with elements of the form $x, y, u$ and $v$. For $a=x_ay_au_av_a\cdots,b=x_by_bu_bv_b\cdots\in\mathfrak{A}=\star_{i\in\{1,2,3,4\}}\mathfrak{A}_i$, define the generalized state $\omega$ by
\begin{align}\label{eq:qss}
\omega(b^*,a)=&\sum_{\ket{\phi}}\overline{\mathcal{A}}(\ket{\phi},x_b\cdots,y_b\cdots,u_b\cdots,v_b\cdots)
\\
&\times\mathcal{A}(\ket{\phi},x_a\cdots,y_a\cdots,u_a\cdots,v_a\cdots),
\end{align}
where the overline denotes complex conjugation, the sum $\sum_{\ket{\phi}}$ is over an orthonormal basis of the state space, and similar to the above example we put the elements into different positions according to their original factors. The defining properties of the generalized state clearly hold. 

\section{The quantum fuzz}\label{sec:tm}

The quantum switch can be generalized into what we call a ``quantum fuzz''. Define
\begin{align}
\mathcal{A}(\ket{\phi},&x,y,u,v)\nonumber
\\
=\bra{\phi}v \Big(&\int_\alpha d\mu(\alpha)\, \ketbra{\alpha}\otimes U_{v,x}^{(\alpha)} x U_{x,y}^{(\alpha)} y U_{y,u}^{(\alpha)}+\nonumber
\\
&\int_\beta d\mu(\beta)\, \ketbra{\beta}\otimes U_{v,y}^{(\beta)} y U_{y,x}^{(\beta)} x U_{x,u}^{(\beta)}\Big)u \ket{\psi}.\label{eq:qfa}
\end{align}
Similar to the quantum switch, for $a,b\in\mathfrak{A}=\star_{i=1,2,3,4}\mathfrak{A}_i$, define the generalized state $\omega$ by
\begin{align}
\omega(b^*,a)=&\sum_{\ket{\phi}}\overline{\mathcal{A}}(\ket{\phi},x_b\cdots,y_b\cdots,u_b\cdots,v_b\cdots)\nonumber
\\
&\times\mathcal{A}(\ket{\phi},x_a\cdots,y_a\cdots,u_a\cdots,v_a\cdots),\label{eq:qfgs}
\end{align}
where similar to the examples above we put the elements into different positions according to their original factors. The defining properties of the generalized state hold, when $d\mu$ preserves the normalization.

The two-point correlation strength for quantum fields depend on the spacetime relation of the two points. In the quantum fuzz we can model different spacetime relations and hence different correlation strengths by choosing the $U^{(\alpha)}$'s and $U^{(\beta)}$'s. The whole construction puts different spacetime relations and correlation strengths in a superposition.

\section{Outlook: Superspacetime}

\begin{figure}
    \centering
    \includegraphics[width=.5\textwidth]{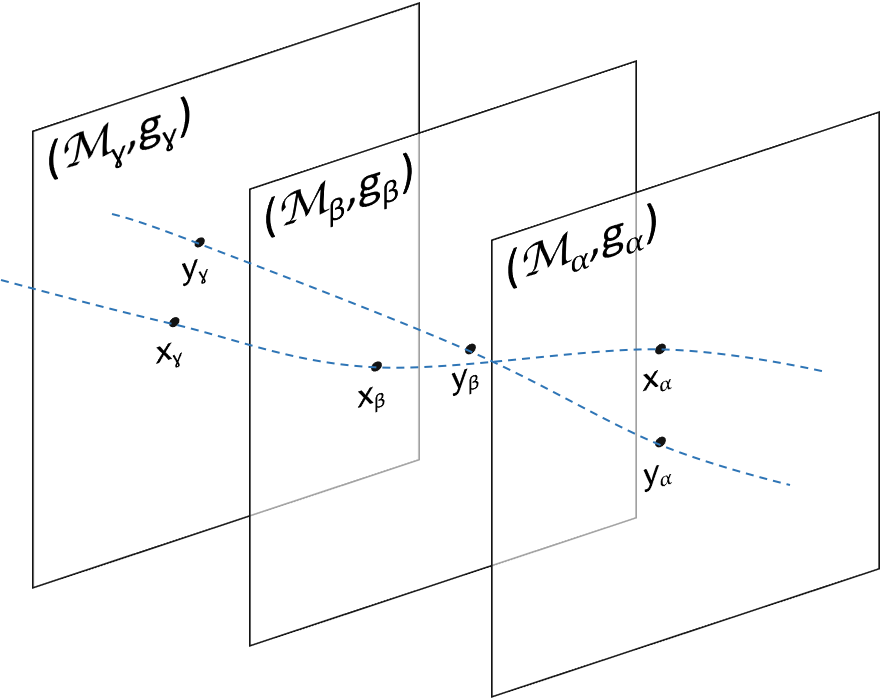}
    \caption{Superspacetime with $x_\alpha,x_\beta,x_\gamma$ identified and $y_\alpha,y_\beta,y_\gamma$ identified. Within different $(\mathcal{M}_\chi,g_\chi)$, $x_\chi$ and $y_\chi$ may have different spacetime relations. Superspacetime puts these relations in a ``superposition''.}
    \label{fig:bp}
\end{figure}

The basic framework presented above already provides enough structures to define and study certain physical concepts (such as generalized entanglement\footnote{See \cite{jia2017generalizing} for a study on finite-dimensional systems.}) and models (such as the quantum switch) that incorporate quantum causality.


To investigate the impact of quantum causality on specific Lagrangian quantum field theories such as QED, we introduce an additional structure called ``superspacetime'', which models the superposition of spacetimes.
A \textbf{superspacetime} $(M,\mathfrak{M},\mathfrak{F})$ is described by a reference set $\mathcal{M}$, a family of spacetimes $\mathfrak{M}=\{(\mathcal{M}_\alpha,g_\alpha)\}_{\alpha\in A}$, and a corresponding family of identification maps $\mathfrak{F}=\{f_\alpha\}_{\alpha\in A}$, where $f_\alpha: M\rightarrow \mathcal{M}_\alpha$ are bijective maps. For each $x\in M$, the points $\{f_\alpha(x)\}_\alpha$ are identified in the superposition of spacetimes.


Physically, this point identification structure constrains the matter field configurations in a functional integral, so that the same field value must be taken on the identified points. For instance, for a real scalar field configuration  $\phi:\prod_\alpha\mathcal{M}_\alpha\rightarrow \mathbb{R}$, $\phi(x_\alpha)=\phi(x_\beta)$ if $x_\alpha$ and $x_\beta$ are identified. In addition, identified points share the same functional form for the Lagrangian density.

Combined with superspacetime, the algebra $\mathfrak{A}$ is localized on $M$ so that each $a\in \mathfrak{A}$ is associated with a subset $P_a$ of $M$. This induces that $a$ is associated with $f_\alpha(P_a)$ on $\mathcal{M}_\alpha$.
$a,b\in \mathfrak{A}$ from different factors are generically in a superposition of spacetime relations, as for different $\alpha$, $f_\alpha(P_a)$ and $f_\alpha(P_b)$ are generically in different spacetime relations. The generalized state, since it now encodes information of the dynamics, depends both on the input ``boundary conditions'' (as in traditional path integrals) and on the Lagrangian. In evaluating the functional integral over superspacetime, a transition amplitude is obtained by summing over the transition amplitudes of all the individual spacetimes $\mathcal{M}_\alpha$.

To illustrate, consider the quantum fuzz defined by (\ref{eq:qfa}) and (\ref{eq:qfgs}). A Lagrangian density (for the matter fields) given on $M$ induces a Lagrangian density on each $\mathcal{M}_\alpha$ and $\mathcal{M}_\beta$ through the identification maps $f_\alpha$ and $f_\beta$. For each $\alpha$ or $\beta$, the functional integral fixes the $U^{(\alpha)}$'s or $U^{(\beta)}$'s as in traditional functional integral on one spacetime. The vector $\ket{\psi}$ contains the information about the matter sector input state and the amplitudes for the spacetimes in superposition. The amplitudes for the spacetime can in principle be calculated from the gravitational Lagrangian given suitable boundary conditions, and ultimately depend on the input data for the quantum spacetime (analogous to the input data of the in- and out-states in the traditional path integral for particle physics). For an optical quantum switch, the states $\ket{\alpha}=\ket{0}$ and $\ket{\beta}=\ket{1}$ are physical systems that evolve in time. In contrast, on superspacetime the states $\alpha$ and $\beta$ represent four-dimensional spacetime configurations that do not evolve in time.

This model provides a good starting point to investigate the impact of spacetime superposition on Lagrangian quantum field theories. Due to the superposition of different spacetime relations one expects that the traditional propagators will be modified. This may render the theory UV-finite \cite{jia2018quantum2}.


\section*{acknowledgements}
I am very grateful to Lucien Hardy, Achim Kempf, Fabio Costa, Matti Raasakka, Rafael Sorkin, and Ognyan Oreshkov for valuable discussions/correspondences.

Research at Perimeter Institute is supported by the Government of Canada through the Department of Innovation, Science and Economic Development Canada and by the Province of Ontario through the Ministry of Research, Innovation and Science. This publication was made possible through the support of a grant from the John Templeton Foundation. The opinions expressed in this publication are those of the authors and do not necessarily reflect the views of the John Templeton Foundation.

\appendix

\section{Hilbert space construction}\label{sec:hs}

The Hilbert space is not fundamental to the present framework, just like it is not to the path integral/functional integral formalism and traditional algebraic formulation of quantum physics. This means that it is not necessary to use the Hilbert space structure to formulate quantum physics in the framework, although a Hilbert space could be introduced for practical purposes.

Here we present one way to construct Hilbert spaces from the basic elements of the framework under an additional assumption, following a procedure similar to the GNS construction \cite{gelfand1943imbedding, segal1947irreducible}. For simplicity we work with $C^*$-algebras rather than general $*$-algebras to avoid having to discuss the domain restrictions of unbounded operators. The construction can be generalized to  $*$-algebras similar to how the ordinary GNS construction can be generalized to to $*$-algebras \cite{khavkine2015algebraic}.

The condition $\omega(a^*,a)\ge 0$ implies two useful properties for $\omega$.
\begin{align}
\omega(a^*,b)=&\overline{\omega(b^*,a)},\label{eq:cs}
\\
\abs{\omega(a^*,b)}^2\le& \omega(a^*,a)\omega(b^*,b).\label{eq:csi}
\end{align}
\begin{proof}
By the definition of generalized states, $\omega((\lambda a+b)^*,\lambda a+b)\ge 0$ for arbitrary $\lambda\in\mathbb{C}, a,b\in \mathfrak{A}$. This means $\abs{\lambda}^2\omega(a^*,a)+\bar{\lambda}\omega(a^*,b)+\lambda\omega(b^*,a)+\omega(b^*,b)$ is a non-negative real number. That the imaginary part vanishes for arbitrary $\lambda$ implies (\ref{eq:cs}). 

(\ref{eq:cs}) in turn implies that the second and third terms sum to $2\Re \lambda\omega(b^*,a)$. If $\omega(a^*,a)=0$, then $2\Re \lambda\omega(b^*,a)+\omega(b^*,b)\ge 0$ for arbitrary $\lambda\in \mathbb{C}$. This implies $\omega(b^*,a)=0$ so (\ref{eq:csi}) holds. If $\omega(a^*,a)> 0$, then pick $\lambda=\overline{\omega(b^*,a)}\mu$, where $\mu\in \mathbb{R}$ is arbitrary. Then $\mu^2\abs{\omega(b^*,a)}^2\omega(a^*,a)+2\mu\abs{\omega(b^*,a)}^2+\omega(b^*,b)\ge 0$. If $\abs{\omega(b^*,a)}=0$, then (\ref{eq:csi}) trivially holds. If $\abs{\omega(b^*,a)}> 0$, then the left hand side is a quadratic polynomial in $\mu$ with a positive leading coefficient, so the discriminant must be non-positive. We have $4\abs{\omega(b^*,a)}^4-4\abs{\omega(b^*,a)}^2\omega(a^*,a)\omega(b^*,b)\le 0$, which implies (\ref{eq:csi}).
\end{proof}

We hope to construct a Hilbert space on $\mathfrak{A}$ as a vector space by taking ``$\braket{a}{b}=\omega(a^*,b)$'' as the inner product. However, this map is truly an inner product only after we mod out the subspace $\mathcal{N}_\omega:=\{a\in \mathfrak{A}:\omega(a^*,a)=0\}$.  ($\mathcal{N}_\omega$ is a subspace, since if $a,b\in\mathcal{N}_\omega$, then $\omega(a^*,b)=\omega(b^*,a)=0$ by (\ref{eq:csi}), whence $\omega((a+b)^*,a+b)=\omega(a^*,b)+\omega(b^*,a)=0$.) 

The map $\braket{\cdot}{\cdot}:\mathfrak{A}/\mathcal{N}_\omega\times \mathfrak{A}/\mathcal{N}_\omega\rightarrow\mathbb{C}, \braket{[a]}{[b]}:=\omega(a^*,b)$ with the equivalence class of $\ket{a}\in\mathfrak{A}$ denoted by $\ket{[a]}$ is well-defined again by (\ref{eq:csi}). A completion in the norm topology yields the Hilbert space $\mathcal{H}_\omega$, with $\ket{\Omega}:=\ket{[e]}$ representing the generalized state $\omega$.

We hope to obtain a representation $\pi_\omega:\mathfrak{A}\rightarrow \mathcal{L}(\mathcal{H}_\omega)$ of actions of the the algebraic elements on the Hilbert space defined on a dense domain $\mathfrak{A}/\mathcal{N}_\omega$ by $\pi_\omega(a)\ket{[b]}:=\ket{[ab]}$. However, in order for this to be well-defined, we need $\mathcal{N}_\omega$ to be a left ideal. This is not true in general but holds for $\mathfrak{A}$ and $\omega$ so that:
\begin{quote}
If $a\in\mathcal{N}_\omega$, then $\omega((ba)^*,ba)=0$ for all $b\in \mathfrak{A}$.
\end{quote}
In contrast to the GNS construction for ordinary QFT, $\pi_\omega$ is not a  *-representation, i.e., $\pi_\omega(a)^*=\pi_\omega(a^*)$ does not hold in general. Nevertheless we obtain a Hilbert space representation for $\omega$ with the expected property
\begin{align}
\omega(a^*,b)=\braket{[ae]}{[be]}=\bra{\Omega}\pi_\omega(a)^*\pi_\omega(b){\ket{\Omega}}.
\end{align}

\bibliography{reference}   

\end{document}